\documentclass[12pt]{article}

\usepackage{amssymb,graphics}

\begin{document}

\begin{center}

{\Large \bf Duality between existence condition and construction
procedure for interface states in a class of narrow-gap
heterostructures \footnote{Talk given at the conference
"Geometry, Integrability and Nonlinearity in Condensed Matter and
Soft Condensed Matter Physics", July 15--20, 2001, Bansko,
Bulgaria}}

\end{center}

\begin{center}

{\bf B. D. Kandilarov $\,^{\dag} $ and V. Detcheva $\,^{\ddag} $}

\end{center}

\begin{center}

$\,^{\dag} $ Institute of Nuclear Research and Nuclear Energy,
72 Tzarigraddsko Chaussee, 1784 Sofia, Bulgaria \\
$\,^{\ddag} $
Department of Condensed Matter Physics,
Sofia University St. Kliment Ochridsky, 1126 Sofia, Bulgaria

\end{center}

\bigskip

\begin{abstract}
 The problem for interface solutions in the
quantum theory of heterostructures comprising narrow-gap semiconductors is
reformulated in the language of commutative diagrams. By this way the theory
of interface states in such heterostructures is naturally factorized in the
 two subproblems: (i) {\it Criterion for the existence of interface
states} , and (ii) {\it Their localization in the common energy gap};
the solution of both of them being presented by the requirement for
commutativity of the relevant diagrams. It is shown that these two problems
are dual in the sense of categorical duality, the passing from the one
commutative diagram to the other being realized by a (contravariant) functor
{\it Op}.

\end{abstract}

\bigskip

\section{Introduction}
\label{intro}

One of the characteristic features of mathematical physics is the
use of abstract notions from mathematics in shedding light on physical
problems and, along with this, the use of commutative diagrams by the
discussion of general concepts in quantum physics [1,2]. However, the
application of the latter technique to particular, even though important
questions from the quantum theory of solids is rather inchoate. Thus, it is
our aim in what follows to sift into the essence of such a particular
problem making use of the language of commutative diagrams. More
specifically, we shall speculate on the interrelation between the criterion
for existence of interface states, on the one hand, and their localization
in the common energy gap at the interface, on the other. To be definite, we
shall consider an abrupt heterostructure comprising two different narrow-gap
semiconductors in epitaxial contact.

In order to make the reading of the present paper as self-consistent as
possible, let us sketch in the physical object under consideration.
Heterostructures between narrow-gap semiconductors belong to the most
important ingredients in solid-state electronics [3], and the two-band
narrow-gap approximation, introduced initially by Keldysh [4] and Wolff [5],
grew up lately in the method of choice by the discussion of their basic
properties. Nowadays, there exists a variety of such approximations, and one
could expect a new revival in this field due to recent experiments [6]
providing strong confirmation of the wave-function hybridizations in
narrow-gap heterostructures.

As a matter of fact, the theory of such heterostructures splits mainly in
two quite different hybridization-based approaches: (i) considerations based
on a Dirac-type equation (for a review see e.g. [7]), and (ii) a
scattering-matrix theory [8,9] based on the hybridization procedure from the
narrow-gap approximation of Pendry and Gurman [10,11] and subsequently
generalized to account for the band-edge discontinuities in the energy-band
profile at the interface [12,13]. It is this latter approach that we are
here using as a starting point in what follows.

\section{Preliminaries}
\label{sec:1}

To begin with, let us recall that a quantum theory of crystalline
semiconductor may be characterized by the pair \{$G$,$\,E_g$\}, $G$ and $E_g$
being the relevant crystal space group and the width of the energy gap,
respectively. In what follows we are dealing with crystals with symmetry
characterized by the presence of a mirror plane and a centre of symmetry in
this plane. As regards the energy gaps, they are taken to be small and for
simplicity, we suppose that they are opened in centre or edge of the
corresponding Brillouin zone. (The generalization to gaps opened in a
general position inside the Brillouin zone is straightforward [11]). By this
way we consider heterostructures comprising two different crystalline
materials of the above mentioned type, the interface being chosen parallel
to the mirror planes characterizing the symmetry of the two crystals in
epitaxial contact. For such heterostructures the $S$-matrix
characterizing the interface as a scatterer
\begin{eqnarray*}\label{eq:sca}
S=\left(\begin{array}{cc}
r & \tau \\ t & \rho
\end{array} \right)
\end{eqnarray*}
takes the form [8]
\begin{eqnarray}\label{eq:scat}
\left(
\begin{array}{cc}
\alpha \cdot e^{i\psi }  &
-i\sqrt{n(1-\alpha ^2)}e^{i(\psi +\varphi )/2} \\
-i\sqrt{{(1-\alpha ^2)\over n}}e^{i(\psi +\varphi )/2} &
\alpha \cdot e^{i\varphi  }
\end{array}
\right)
\end{eqnarray}
where $n$ is defined by the ratio of the group velocities of the
flux-carrying Bloch waves on both sides of the interface; the real-valued
positive quantities $\alpha $ and $\psi $ stand for the modulus and the
argument of the reflection coefficient $r$, respectively; and the real
positive $\varphi $ is the argument of the other reflection coefficient $%
\rho $. The main result thus obtained in [8] is that in this case there may
exists at most one interface state, the necessary and sufficient condition
for its appearance being given by
\begin{equation}
\tan \left({\psi +\phi _1  \over 2}
\right)={1-\alpha \over 1+\alpha }\cot \left({ \varphi
+\phi _2\over 2}\right).
\end{equation}
Here $\phi _j$ $(j=1,2)$ are parameters characterizing the energy gap of the
left $(j=1)$ and the right $(j=2)$ crystals.

\section{Two rigorous results and their physical interpretation}
\label{2}

The problem that faces us now is how to factorize the physical
interpretation of $(2)$ into two rigorously defined subproblems: (i) for
existence of an interface solution and, if such a localized state does
exist, (ii) for finding its position in the common energy gap. The
reformulation of these problems in the language of commutative diagrams
leads in a natural way to the nontrivial result that they are dual in the
sense of categorical duality\footnote{For the mathematical tools from the
theory of commutative diagrams, needed for what follows, see e.g. [14].}.
However, the price of this is that in what follows we are forced to use
a kind of mathematical manner, typical for the exposition of such
results.  Hence, both of the above subproblems have to be formulated as
relevant {\it propositions}, followed by the rigorous {\it proof}  of each
of them.

As regards e.g. the first subproblem, its physical essence reduces to
traversing the common energy gap in search for such values of energy, for
which eq. (2) holds. Thus rigorously speaking, we arrive at our

\textbf{Proposition 1}. \textit{The criterion for existence of interface
states is given by the requirement for the diagram }
\begin{equation}
\begin{array}{lll}
\begin{array}{l}
\\
\,\,E(\in E_g)
\end{array}
&
\begin{array}{l}
\\
\stackrel{p_2}{\longrightarrow }
\end{array}
&
\begin{array}{l}
\, \\
\,\,\,\,\,\,\,\,\,\,\,\,\,\,\,\phi _2
\end{array}
\\
\begin{array}{l}
\,\,\,\,\,\,\,\,\, \\
p_1\downarrow
\end{array}
&  &
\begin{array}{l}
\,\,\,\,\,\,\,\,\,\,\, \\
\,\,\,\,\,\,\,\,\,\,\downarrow q_2
\end{array}
\\
\begin{array}{l}
\,\,\, \\
\,\,\,\,\,\,\,\phi _1
\end{array}
&\stackrel{q_1}{\longrightarrow }
 &
\begin{array}{l}
\\
I_1=I_2=I_0
\end{array}
\end{array}
\end{equation}
{\it to be commutative, i.e.}
\begin{equation}
q_1\circ p_1=q_2\circ p_2
\end{equation}
\textit{for some }$E\in E_g$\textit{, for which }$E=dom(q_1\circ
p_1)=dom(q_2\circ p_2)$\textit{\ is the energy of the interface state}.

Let us only sketch the

\textbf{Proof}: To every value of the energy from the
common gap $E_g$ we juxtapose a corresponding value of $\phi _j$ ($j=1,2$)
by the arrows
\begin{equation}
p_j: E \longmapsto \phi _j
\qquad j=1,2 .
\end{equation}
Then, defining the arrows $q_j\,\,$($j=1,2$) by
\begin{eqnarray}
q_1 &:&\phi _1\longmapsto I_1=\tan \left({\psi +\phi _1\over 2}\right)
\\
q_2 &:&\phi _2\longmapsto I_2={1-\alpha \over 1+\alpha }\cot
\left({\varphi +\phi _2\over 2} \right),
\nonumber
\end{eqnarray}
we construct the composites of arrows $q_1\circ p_1$ and $q_2\circ
p_2$. Traversing the common energy gap in search for such values of $E\in
E_g $, for which the diagram is commutative, we easily observe that this is
just this $E\equiv E_{is}$ which is obtained as solution of (2).
$\blacksquare $

The particular value of $I_0(=I_1=I_2)$ thus obtained is the starting point
by the discussion of the second subproblem we are here interested in,
namely, the localization of the interface state in the common gap.
Obviously, in order to answer this question it is sufficient that at least
one of the compositions $p_1^{-1}\circ q_1^{-1}$ or $p_2^{-1}\circ q_2^{-1}$
do exist. As a matter of fact, for this particular case we shall prove a
more general result in the next

\textbf{Proposition 2. }\textit{To any commutative diagram (of the category
of diagrams (3))defining the existence of an interface state corresponds a
dual to it (''opposite'') diagram which specifies the position of this
interface state in the common energy gap.}

\textbf{Proof}: The duality principle is a handy way to have (at once) the
dual theorem and, in the categorical language, duality is defined by the
process ''Reverse all arrows'' [14]. Thus, the diagram dual to our diagram
(3) comprises the composite arrows $(q_jp_j)^{-1}=p_j^{-1}\circ q_j^{-1}\,$($%
j=1,2$). On account of (5) immediately follow
\begin{eqnarray}
q_1^{-1} &:&I_0\longmapsto \phi _1=2\cdot \tan ^{-1}I_0-\psi   \\
q_2^{-1} &:&I_0\longmapsto \phi _2=2\cdot \cot ^{-1}\left(I_0{1+\alpha
\over 1-\alpha}\right) )-\varphi \,.  \nonumber
\end{eqnarray}
What remains to be done is to deduce the explicit form of the arrows $%
p_j^{-1} $ ($j=1,2$). From the hybridization procedure for two-band
narrow-gap semiconductors [10,11] one obtains
\begin{equation}
{v_j^{in}(K_j-K_{0j})-(E-E_{0j})\over v_j^{in}}+(V_j^{\prime
})_{+-}\cdot e^{i\phi _j}=0, \quad   j=1,2,
\end{equation}
where $K_j$ is the component of the Bloch-wave vector (for the
relevant crystal) which is perpendicular to the interface; $E_{oj}$ and $%
K_{0j}$ designate the point in the Brillouin zone of the $j-$th crystal
where the energy gap is opened; the quantity
\[
(V_j^{\prime })_{+-}=\langle a_j^{+}\mid V_j^{\prime }\mid a_j^{-}\rangle
\]

\noindent is the matrix element of this component $V_j^{\prime }$ of the
relevant crystal potential which is responsible for opening the
corresponding energy gap of the $j-$th crystal by the narrow-gap
hybridization with the Bloch waves $a_j^{\pm }$; and $v_j^{in}$ are the
group velocities of the incoming waves on both sides of the interface.

For the particular case of an energy gap opened in centre or edge of the
Brillouin zone, $K_j$ is complex-valued, its real part being
$Re(K_j)=K_{0j}$.
Thus, we obtain from (8)
\begin{equation}
 p_j^{-1}:\phi _j\longmapsto E=E_{0j}+v_j^{in}(V_j^{\prime })_{+-}\cos
\phi _j, \qquad  j=1,2.
\end{equation}
From (7) $\div $ (9) immediately follows that for the case under
consideration both composite arrows $p_j^{-1}\circ q_j^{-1}$ ($j=1,2$) do
exist and, consequently, the position of the interface state $E_{is}$ in the
common gap at the interface is defined in an unique way by
\begin{equation}
E_{is}=\mbox{codom}\,I_0
\end{equation}
This completes the proof. $\blacksquare $

\textbf{Corollary 1}. {\it Making use of e.g. the composite arrow
$p_1^{-1}\circ q_1^{-1}$ from (10) we obtain}
\begin{equation}
E_{is}=E_{01}+v_1^{in}(V_1^{\prime })_{+-}\cdot \cos (2\tan ^{-1}I_0-\psi
).
\end{equation}

\textbf{Corollary 2}. {\it From the duality of the two problems - for
existence of an interface state and for finding its position in the common
energy gap - and as a result of their reformulation in terms of mutually
opposite diagrams immediately follows that the passage from the one to the
other is obtain under the action of a contravariant functor $Op$}.

\section{Conclusions}\label{sec:3}

We close with some remarks concerning the comparison between the present
approach with the other one (based on a Dirac-type equation) we have
referred to in the beginning.

Let us first of all mention that the name 'Dirac-type equation' is somewhat
misleading for this is only a suggestion that it is \textit{formally}
obtained from the relativistic wave equation of Dirac by replacing the
universal constant 'velocity of light $c$' by the so called 'interband
velocity matrix element $v$' [15]. Hence, the relevant theory of such
interface states has nothing to do with the relativistic-quantum theory of
interface states ([16], see also [17,18]).

In addition, it is worth noting that the present alternative approach seems
to be more convenient by the discussion of the essence of basic problems in
the two-band narrow-gap quantum theory of semiconductor heterostructures.

An instructive example in this direction is the duality between
existence condition and construction procedure for the particular
above discussed problem. It may be also considered as a nice
example for unusual but useful application of a modern
mathematical tool-- the notion of of commutative diagrams-- to a
particular problem from the theory of solids. It is worth nothing
that, as regards the scattering-theoretical approach to
heterojunction problems, it is important to take into account the
{\it specific} features characterizing the scattering by
potentials with {\it different} finite asymptotics for $x \to
-\infty  $ and $x\to +\infty  $, respectively \cite{19,20}. What
is more, as regards the use of heterojunction $S $- matrices
(like those from \cite{8}) as a starting point, the situation is
even additionally complicated due to the need to take into
account Bloch waves in the  two {\it different} crystals in
epitaxial contact and the relevant generalizations of the
properties characterizing a {\it heterojunction} $S $- matrix.

\section*{Acknowledgements} Thanks are due to R. Rajaraman and G.
Grahovski for useful discussions.

\end{document}